\documentclass[11pt,preprint]{aastex}

\usepackage{graphics,graphicx}
\usepackage{natbib}
\usepackage{float}
\def\apjl{\rm ApJL}
\def\apjs{\rm ApJS}

\def\aap{\rm AAP}

\newcommand{\gtorder}{\mathrel{\raise.3ex\hbox{$>$}\mkern-14mu
            \lower0.6ex\hbox{$\sim$}}}
\newcommand{\ltorder}{\mathrel{\raise.3ex\hbox{$<$}\mkern-14mu
            \lower0.6ex\hbox{$\sim$}}}

\shorttitle{Quasi-periodic oscillations from SGR 1806--20}
\shortauthors{Miller, Chirenti, and Strohmayer}

\begin{document}

\title{ON THE PERSISTENCE OF QPOs DURING THE SGR 1806$-$20 GIANT FLARE}

\author{M. Coleman Miller\altaffilmark{1}, Cecilia Chirenti\altaffilmark{2}, and Tod E. Strohmayer\altaffilmark{3}}

\affil{
{$^1$}{Department of Astronomy and Joint Space-Science Institute, University of Maryland, College Park, MD 20742-2421, USA}\\
{$^2$}{Centro de Matem\'{a}tica, Computa\c{c}ao e Cogni\c{c}ao, UFABC, 09210-170 Santo Andr\'{e}-SP, Brazil}\\
{$^3$}{Astrophysics Science Division and Joint Space-Science Institute, NASA’s Goddard Space Flight Center, Greenbelt, MD 20771, USA}\\
}

\begin{abstract}

The discovery of quasi-periodic brightness oscillations (QPOs) in the X-ray emission accompanying the giant flares of the soft gamma-ray repeaters SGR~1806--20 and SGR~1900+14 has led to intense speculation about their nature and what they might reveal about the interiors of neutron stars.  Here we take a fresh look at the giant flare data for SGR~1806--20, and in particular we analyze short segments of the post-peak emission using a Bayesian procedure that has not previously been applied to these data.  We find at best weak evidence that any QPO persists for more than $\sim 1$ second; instead, almost all the data are consistent with a picture in which there are numerous independently-excited modes that decay within a few tenths of a second.  This has interesting implications for the rapidity of decay of the QPO modes, which could occur by the previously-suggested mechanism of coupling to the MHD continuum.  The strongest QPOs favor certain rotational phases, which might suggest special regions of the crust or of the magnetosphere.  We also find several previously unreported QPOs in these data, which may help in tracking down their origin.

\end{abstract}


\section{INTRODUCTION}
\label{sec:introduction}

Quasi-periodic X-ray brightness oscillations in the 2004 December giant flare from SGR 1806$-$20 were first reported by \citet{2005ApJ...628L..53I}. They detected a strong QPO at 92.5~Hz in Rossi X-Ray Timing Explorer (RXTE) data and found evidence for lower frequency QPOs at about 18 and 30~Hz.  They suggested the QPOs could be related to torsional modes excited during the flare.  Motivated by this \citet{2005ApJ...632L.111S} investigated the RXTE data obtained during the 1998 August event from SGR 1900+14.  They found a sequence of rotation-phase-dependent QPOs at 28, 54, 84 and 155~Hz, and suggested an identification with a sequence of low-order torsional modes with different $l$ values.  These authors also re-examined the RXTE data from the SGR 1806$-$20 giant flare, and found evidence for additional oscillations at 150 and 625 Hz \citep{2006ApJ...653..593S}.  They suggested that the 625~Hz QPO could be identified with torsional oscillation modes with at least one radial node in the crust, and that this could lead to a probe of the crust thickness.

These observational findings touched off a plethora of theoretical investigations.  \citet{2006MNRAS.368L..35L} noted that the coupling of the crust to the core by a strong vertical magnetic field should lead to damping of the crustal oscillations within a few tenths of a second. Subsequent investigations explored the magnetic coupling of the crust to the core in more realistic scenarios \citep{2007MNRAS.377..159L,2007Ap&SS.308..607G,2007ApJ...661.1089V,2009MNRAS.396.1441C,2011MNRAS.418..659L,2011MNRAS.410L..37G,2013MNRAS.430.1811G}, and also explored how torsional mode identifications could be used to constrain the properties of the neutron star, including its mass and radius \citep{2007MNRAS.374..256S,2007MNRAS.375..261S,2011MNRAS.414.3014C,2016NewA...43...80S,2016EPJA...52...63M,2017MNRAS.464.3101S}, as well as its internal composition \citep{2007MNRAS.379L..63W}.  Other authors have explored emission and X-ray modulation mechanisms for the QPOs \citep{2008ApJ...680.1398T,2012ApJ...751L..41D}, and the role of superfluidity in modifying the torsional oscillation spectrum of neutron stars \citep{2013MNRAS.429..767P,2014MNRAS.438..156P}.  A focus of current theoretical modeling is the relative importance of magnetic and shear stresses, and to what extent the crustal torsional oscillation spectrum is present in the excitation spectrum, if at all \citep{2014AN....335..240G,2016MNRAS.460.4242G,2016MNRAS.463.1173P,2016ApJ...823L...1L,2016ApJS..224....6L,2016MNRAS.463.1173P,2016MNRAS.459.4144R,2017PhRvC..95a5803T,2017AN....338.1105G}.

Most of the data analysis and theoretical attention has focused on the frequencies of the QPOs (for a recent example, see \citealt{2018A&A...610A..61P}).  The attraction of this focus is that, potentially, particular oscillation modes can be associated with particular QPO frequencies, and those associations can reveal aspects of the interiors of the oscillating neutron stars.  But the {\it duration} of the signals also contains useful information, particularly about the expected damping mentioned above.  Indeed, \citet{2014ApJ...793..129H} found evidence of such damping of the $\sim 625$~Hz QPO from SGR~1806$-$20.  However, there has not yet been a systematic study of these QPOs over short intervals.

Here we perform a comprehensive analysis of QPOs from the SGR~1806$-$20 giant flare.  We divide the data into intervals of one second, which is roughly one eighth of the rotation period, to explore whether there are short-lived QPOs, or whether instead most QPOs are long-lived.  This also allows us to determine whether there are QPOs that are present for many rotational periods, but only in some range of rotational phases.  Somewhat surprisingly, we find that there is little evidence for long-lived QPOs; instead, there are many QPOs that come and go over times that are likely to be a few tenths of a second.  This could provide further support for the idea that coupling of crustal modes to the core MHD continuum damps the modes rapidly.  In Section~\ref{sec:methods} we describe the data set that we analyze, as well as our analysis method (which is a new Bayesian approach not previously used in this context).  In Section~\ref{sec:results} we present our results, and we conclude with a discussion of the results in Section~\ref{sec:discussion}.

\section{METHODS}
\label{sec:methods}

\subsection{Observations}

The data we analyze are the same as were analyzed by \citet{2005ApJ...628L..53I} and \citet{2006ApJ...653..593S}.  The data were recorded in the Goodxenon\_2s mode, which has a time resolution of $\approx 1~\mu$s and a buffering time of 2 seconds and which thus does not saturate as easily as the Goodxenon\_16s mode that happened to be in use during the earlier giant flare from SGR~1900+14.  We include all counts without making cuts on the energy channel; a motivation for this is that the burst actually came in through the side of the detector, which made particular energy channel assignments less reliable than normal.

\subsection{Data analysis}

\begin{figure}
\begin{center}
  \resizebox{1.0\textwidth}{!}{\includegraphics{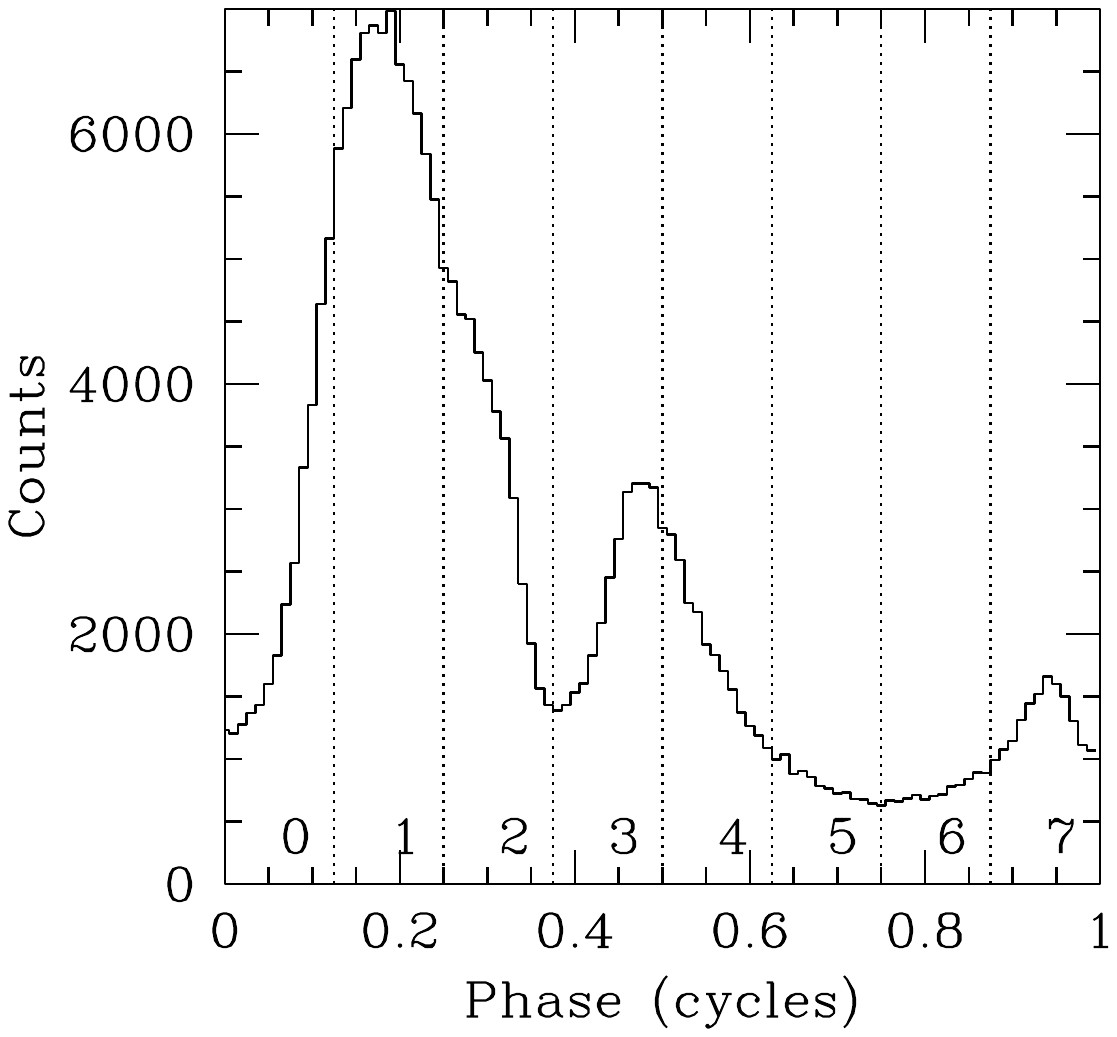}}
\vspace{-2.0truein}
   \caption{Light curve for one rotational period, produced by stacking the first ten periods after the starting point (2004 December 27 at 21:30:51.378 UTC).  The segments labeled 0 through 7 correspond to the segments that we analyze, i.e., we divide each of the 46 rotational periods in our data into eight parts, and analyze 1.0 seconds from the beginning of each segment.  Because 1/8 of a rotational period is 0.945 seconds, this means that each segment overlaps the next by 0.055 seconds; we make this choice because unlike for a 0.945 second segment, a fast Fourier transform of a 1.0 second segment is commensurate with the intrinsic $2^{-20}$ second time binning of the {\it RXTE} data.}
\label{fig:lightcurve}
\end{center}
\end{figure}

The rotational period of SGR~1806$-$20 is 7.56 seconds (e.g., \citealt{2005Natur.434.1098H}).  In order to keep rotational phase information it would be useful to analyze data in segments with durations that are evenly divisible into the rotational period; for example, 1/8 of a period has a duration of 0.945 seconds.  However, because the {\it RXTE} data in this mode have time bins of duration $2^{-20}$ seconds, a fast Fourier transform of data from 0.945-second segments risks aliasing because 0.945 seconds is not a power of 2 times $2^{-20}$ seconds.  We therefore divide data into segments that start every 0.945 seconds, but that are exactly 1 second long each; thus segments overlap each other by 0.055 segments.  We have verified via spot checks that using 0.945-second segments or 1.0-second segments produces results that are qualitatively and quantitatively similar to each other.  There are 368 such 1-second segments in the data, lasting a total of $\approx 348$ seconds.  We also explored whether a sliding window would reveal additional features, but the results were all qualitatively the same as those we obtained in the analyses that we report.  We use data starting 20 seconds after the beginning of the burst (that is, our start time is 2004 December 27 at 21:30:51.378 UTC) to avoid saturation of the {\it RXTE} Proportional Counter Array detectors.  A stacked light curve for the eight segments we use, based on the first ten periods after our starting point, is in Figure~\ref{fig:lightcurve}; compare this figure with the bottom panel of Figure~1 from \citet{2006ApJ...653..593S}, and note that the top panel of their same figure shows the full light curve of the giant flare including the obvious rotational modulation.

For each segment, we divide our data into 4096 equal-length bins in time, which therefore last exactly $2^{-12}$ seconds each, and use the binned data to construct a power density spectrum from a fast Fourier transform.  The resulting spectrum for each segment therefore extends to a Nyquist frequency of 2048~Hz in intervals of 1~Hz.  Instead of the commonly-used Leahy normalization of the power density spectrum \citep{1983ApJ...266..160L}, in which the mean power for pure Poisson noise is 2, we use a normalization in which the mean for Poisson noise is 1, to match the assumption of \citet{1975ApJS...29..285G}, which we follow in our analysis of the power density spectrum.  Our analysis does not consider frequencies below 10~Hz, because lower frequencies have greater red noise contributions from the pulse profile and the overall decay of the emission.  We need to consider frequencies at least this low given the reports of some QPOs in the $10-20$~Hz range.  The contribution of the red noise in the vicinity of 10~Hz is small, which means that our conclusions are insensitive to the precise choice of the frequency floor.

Our Bayesian search for QPOs compares two models:

\begin{enumerate}

\item Red noise model: there is a red noise component, so that the non-Poisson power density at frequency $f$ is 
\begin{equation}
P_{\rm red}(f)=A(f/15~{\rm Hz})^{-2}\; ,
\end{equation}
where we normalize the power to a fiducial frequency of $15~{\rm Hz}$.  Thus the amplitude $A$ is the only parameter in this model.  We chose an $f^{-2}$ slope for our red noise model because this would correspond to an exponential decay.  However, we found that the choice of the red noise slope makes no significant difference. In particular, we reanalyzed the strongest signals (listed in Table~\ref{tab:QPOs} below) using red noise power laws of $f^{-1}$, $f^{-1.5}$, $f^{-2.5}$, and $f^{-3.0}$ for both the red noise and Lorentzian models, and all of the Bayes factors were consistent with the values that we report here.  We have also carried out an exploratory analysis in which we include the red noise slope as an extra parameter, and the results are generally consistent with the $f^{-2}$ red noise fit.

\item Lorentzian model: there is a red noise component plus a Lorentzian, so that the non-Poisson power density at frequency $f$ is
\begin{equation}
P_{\rm Lorentz}(f)=B(f/15~{\rm Hz})^{-2}+{C\over{(\Delta f)^2+(f-f_0)^2}}\; .
\label{eq:Lorentz}
\end{equation}
Thus this model has four parameters: $B, C, \Delta f, {\rm and}~f_0$.  The red noise amplitude $B$ for the Lorentzian model is determined independently of the parameter $A$ in the red noise model.

\end{enumerate}

We fit each of these two models to all of the 368 data segments, independently.  That is, we allow the parameters of the models to vary freely from one data segment to the next.  We follow standard Bayesian procedure by computing an unnormalized posterior probability density $p$ at a given parameter combination by multiplying the prior probability density $q$ at that combination by the likelihood ${\cal L}$ of the data given the model at that combination.  Thus for the Lorentzian model,
\begin{equation}
p_{\rm Lorentz}(B,C,\Delta f,f_0)\propto q_{\rm Lorentz}(B,C,\Delta f,f_0){\cal L}_{\rm Lorentz}(B,C,\Delta f,f_0)\; .
\end{equation}
The priors are normalized so that their integral over the parameter space is 1, whereas the posteriors would have to be divided by their integral over all parameter space in order to be normalized.

We compute the likelihood using the formulae of \citet{1975ApJS...29..285G}, who showed that, for a single frequency bin ($n=1$ in his equation~(15)), the likelihood of observing a power $P$ when a non-Poisson power $P_s$ is expected (and when the power is normalized such that the mean Poisson-only power is 1) is
\begin{equation}
{\cal L}(P;P_s)=e^{-(P+P_s)}\sum_{m=0}^\infty {P^m P_s^m\over{(m!)^2}}\; .
\end{equation}
In our analysis we use this formalism along with the standard assumption that the amplitudes at different frequencies in the fast Fourier transform are statistically independent from each other.  Note that in the limit of no non-Poisson power, i.e., $P_s\rightarrow 0$, only $m=0$ contributes in the sum and thus ${\cal L}(P;0)\rightarrow e^{-P}$, which is the familiar result for Poisson-only power with a mean of 1.

The likelihood of the entire power spectrum that we analyze for a given segment, which has 2038 independent frequencies (because 10 of the 2048 total frequencies are below 10~Hz), is the product of the likelihoods of the observed power at each frequency.  For example, for the Lorentzian model,
\begin{equation}
{\cal L}_{\rm Lorentz}(B,C,\Delta f,f_0)=\prod_i \left[e^{-[P_i+P_{\rm Lorentz}(f_i|B,C,\Delta f,f_0)]}\sum_{m=0}^\infty {P_i^m P^m_{\rm Lorentz}(f_i|B,C,\Delta f,f_0)\over{(m!)^2}}\right]\; .
\end{equation}
Here $P_i$ is the observed power at the frequency $f_i$ of bin $i$, and $P_{\rm Lorentz}(f_i|B,C,\Delta f,f_0)$ is the non-Poisson power expected at frequency $f_i$ given model parameter values $B,C,\Delta f$, and $f_0$.  We approximate the infinite sum at each frequency by truncating the series when the term being considered has a magnitude less than $10^{-20}$ of the running sum, which we find gives a fast and accurate measure of the likelihood.

Our prior for the red noise amplitude, for both models, is uniform between 0 and 10 at 15~Hz.  Thus the red noise component is $[0-10](f/15~{\rm Hz})^{-2}$; recall that we are fitting power densities, so for instance 0 would mean no non-Poisson noise.  For no segment does the best-fit red noise amplitude approach 10, and by definition the amplitude cannot be negative.  For similar reasons, our prior for the Lorentzian amplitude $C$ is uniform between 0 and 30.

Our prior for the Lorentzian width $\Delta f$ is uniform in log width, from $\log_{10}(\Delta f/{\rm Hz})=-0.3$ ($\Delta f\approx 0.5$~Hz) to $\log_{10}(\Delta f/{\rm Hz})=2$ ($\Delta f=100$~Hz).  We choose a logarithmic prior so that we are not biased in the scale of the width, and we choose this minimum width because at our frequency resolution of $\sim 1$~Hz, narrower peaks could simply represent single-frequency fluctuations.  Our maximum width encompasses the possibility that there could be an overall excess of noise, rather than a sharp peak.

Our prior for the Lorentzian centroid frequency $f_0$ depends on the search.  We select this approach because it is possible that there are several QPOs present in a given data segment; if we had instead searched for a single Lorentzian over all frequencies, it is probable that only the strongest QPO would have been detected.  In a given search, we assign a prior probability to $f_0$ that is uniform in log frequency within the bounds of that search.  We base our frequency searches on previously reported QPOs, and thus our frequency bounds are $10-20$~Hz; $20-40$~Hz; $40-80$~Hz; $80-120$~Hz; $120-200$~Hz; $200-550$~Hz; $550-700$~Hz; $700-1000$~Hz; $1000-1400$~Hz; and $1400-2000$ Hz.  We find that there are a few data segments that contain multiple strong QPOs, but that this is rare.  We therefore also performed an exploratory analysis of the data segments in which we used the full frequency range $10-2000$~Hz, which led to the same conclusions that we present in this paper.

In order to provide a framework for our later calculation of the significance of possible QPO signals, we also produce a synthetic data set with no red noise and no QPOs.  We do this by generating each power density in the spectrum by random draws from the $e^{-P}$ distribution expected for pure Poisson noise; we found that other methods (for example, generating and Fourier-analyzing a count rate curve that is similar to the real data but that has no periodic signals) produce comparable results.  We then analyze our synthetic set in the same way that we analyze the real data.

We explore the parameter spaces of both models using an affine-invariant Markov chain Monte Carlo (MCMC) code that we wrote based on the approach of \citet{2010CAMCS...5...65G}.  For both the red noise and the Lorentzian search, for every data segment, we perform 20 independent runs with 32 walkers each, and 100 candidate updates of each walker after convergence was established.  We find this to be sufficient in all analyses.  

After the parameter spaces of both models are explored for a given data segment, we estimate the significance of any possible QPOs by computing Bayes factors.  In general, the Bayes factor in favor of some model A (with parameters ${\vec\alpha}$) over some model $B$ (with parameters ${\vec\beta}$) is the ratio
\begin{equation}
{\cal B}_{AB}={\int q_A({\vec\alpha}){\cal L}_A({\vec\alpha})d{\vec\alpha}\over{\int q_B({\vec\beta}){\cal L}_B({\vec\beta})d{\vec\beta}}}\; .
\end{equation}
The integral in the numerator, which is over all possible values of the parameters ${\vec\alpha}$, is sometimes called the {\it evidence} for model A, and likewise the integral in the denominator is the evidence for model B.  The odds ratio ${\cal O}_{AB}$ of model A in favor of model B is simply ${\cal B}_{AB}$ multiplied by the prior probability ratio between model A and model B.  We set that prior probability ratio to unity between the Lorentzian and the red noise model, so ${\cal O}_{\rm Lorentz,red~noise}={\cal B}_{\rm Lorentz,red~noise}$ in our case.

Note that the Bayes factor depends on the priors as well as on the likelihood.  For example, if we were to choose a very narrow prior on $\log_{10} f_0$ that happened to be near a strong signal, the Bayes factor would be much larger than if we chose a broad prior.  This is why we perform the same analysis, using the same ten frequency ranges, on the synthetic data (which has no signal) as we do on the real data.  

\section{RESULTS}
\label{sec:results}

\subsection{General characteristics of the signals}

\begin{figure}
\begin{center}
\vspace{-1.0truein}
\begin{center}
  \resizebox{1.0\textwidth}{!}{\includegraphics{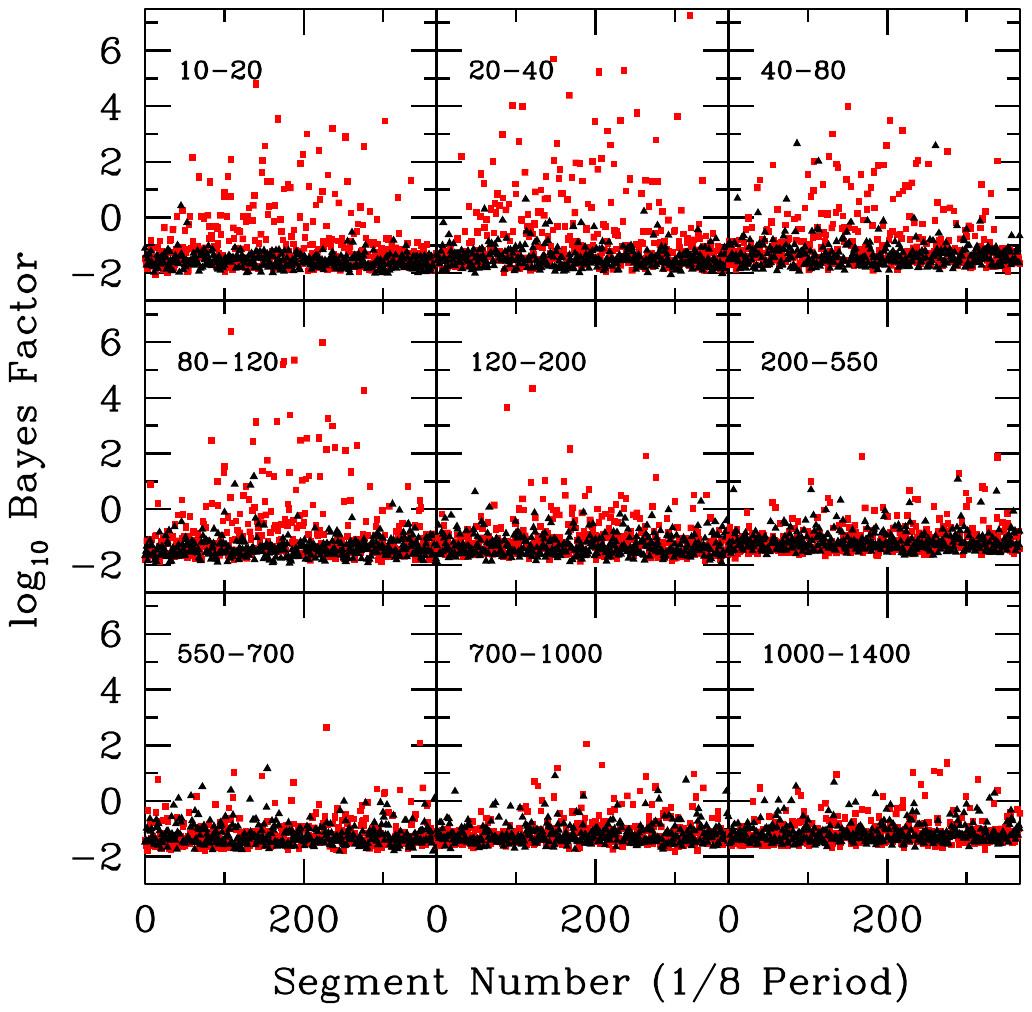}}
\end{center}
\vspace{-2.0truein}
   \caption{Bayes factors, as a function of segment, for the real data (solid red squares) and for the synthetic data (solid black triangles).  Recall that the segments start 1/8 of a period apart, or 0.945 seconds, but last 1.0 seconds each.  The different panels are for centroid frequencies in various ranges, which are given in Hertz in the top left of each panel.  Note that the power floors for the real and the synthetic data are comparable, which suggests that there is no overall bias in the Bayes factors.  Note also that there are far more segments in the real data than in the synthetic data that have Bayes factors greater than unity, or 10, or 100, or other thresholds.  Most of these likely contain real signals, but at small Bayes factors individual claims to reality are highly uncertain.}
\label{fig:bayes9panel}
\end{center}
\end{figure}

Figure~\ref{fig:bayes9panel} shows the Bayes factors in favor of the Lorentzian model as computed for both the real data (solid red squares and lines) and the synthetic data.  There are several points to note:

\begin{enumerate}

\item Consistent with previous analyses, there are far more large Bayes factors in the real data than in the synthetic (no-signal) data.  The highest Bayes factor in the synthetic data is ${\cal B}\approx 450$.  In contrast, there are a large number of segments and frequency ranges in the real data with ${\cal B}>10^3$, with the highest Bayes factor exceeding $10^7$.  

\item Virtually all of the signals last for only one of the 1-second segments.  This suggests that almost all the signals damp rapidly.  See Section~3.2 for an exploration of whether some signals last for multiple periods.

\item In addition to many signals that have been reported previously, there are other frequencies with strong to nearly conclusive significance.  In Table~\ref{tab:QPOs} we list the signals found in our analysis with Bayes factors greater than 1000, i.e., larger than the highest Bayes factor found in the synthetic data, which have best-fit frequency widths less than 10~Hz at 84\% credibility (so that these are likely to be QPOs rather than excess noise over a broad band).  Many of these signals do not appear to have been reported earlier, e.g., the QPOs at 51~Hz, 97~Hz, and 157~Hz. 

\end{enumerate}

\begin{deluxetable}{rrrr}
\tablecaption{QPOs with Bayes factors larger than 1000 and $\Delta f<10$~Hz\label{tab:QPOs}}
%\tabletypesize{\normal}
\tablecolumns{4}
\tablewidth{0pt}
\tablehead{
\colhead{Start time (s)$^a$} & \colhead{Frequency (Hz)$^b$} & \colhead{Freq width (Hz)} & \colhead{Bayes factor}
}
\renewcommand{\baselinestretch}{1.1}
\startdata
301.455 &  $31.499-32.724$ & $1.324-2.640$ & $1.80\times 10^7$\\
102.060 &  $91.239-92.036$ & $0.748-1.593$ & $2.47\times 10^6$\\
210.735 &  $92.967-93.465$ & $0.566-1.066$ & $9.77\times 10^5$\\
138.915 &  $21.148-21.605$ & $0.610-1.097$ & $5.09\times 10^5$\\
177.660 &  $87.870-91.119$ & $2.787-7.296$ & $2.30\times 10^5$\\
165.375 &  $92.109-92.792$ & $0.631-1.324$ & $2.09\times 10^5$\\
223.020 &  $23.302-25.214$ & $1.957-4.915$ & $1.97\times 10^5$\\
192.780 &  $27.574-28.662$ & $0.913-1.956$ & $1.68\times 10^5$\\
{\bf 164.430}$^c$ & {\bf 94.731 -- 100.485} & {\bf 2.380 -- 7.281} & {\bf 1.60}$\times 10^5$\\
132.300 &  $16.621-17.052$ & $0.556-0.893$ & $6.28\times 10^4$\\
{\bf 113.400} & {\bf 156.103 -- 157.109} & {\bf 0.781 -- 1.916} & {\bf 2.21}$\times 10^4$\\
 89.775 &  $25.857-27.221$ & $1.273-2.820$ & $1.10\times 10^4$\\
102.060 &  $22.908-23.506$ & $0.557-0.961$ & $9.76\times 10^3$\\
238.140 &  $28.831-32.546$ & $3.148-9.308$ & $5.63\times 10^3$\\
 83.160 &  $148.988-149.803$ & $0.621-1.172$ & $4.57\times 10^3$\\
286.335 &  $20.202-21.262$ & $1.702-4.525$ & $4.30\times 10^3$\\
218.295 &  $27.267-28.475$ & $0.775-1.643$ & $3.09\times 10^3$\\
172.935 &  $92.024-93.510$ & $0.702-2.698$ & $2.49\times 10^3$\\
218.295 &  $94.308-97.290$ & $2.000-6.738$ & $1.83\times 10^3$\\
156.870 &  $91.605-92.506$ & $0.604-1.159$ & $1.48\times 10^3$\\
132.300 &  $88.422-89.257$ & $0.615-1.380$ & $1.36\times 10^3$\\
203.175 &  $30.329-31.999$ & $1.077-3.396$ & $1.32\times 10^3$\\
{\bf 123.795} &  {\bf 50.962 -- 51.984} & {\bf 0.641 -- 3.249} & {\bf 1.02}$\times 10^3$\\
\enddata
\tablenotetext{a}{$t=0$ corresponds to 2004 December 27 at 21:30:51.378 UTC.}
\tablenotetext{b}{The numbers for the frequency and frequency width show the 16\% to 84\% credibility range.}
\tablenotetext{c}{Boldface indicates a frequency not previously reported.}
\end{deluxetable}

\renewcommand{\baselinestretch}{1.0}

\begin{figure}
\begin{center}
  \resizebox{1.0\textwidth}{!}{\includegraphics{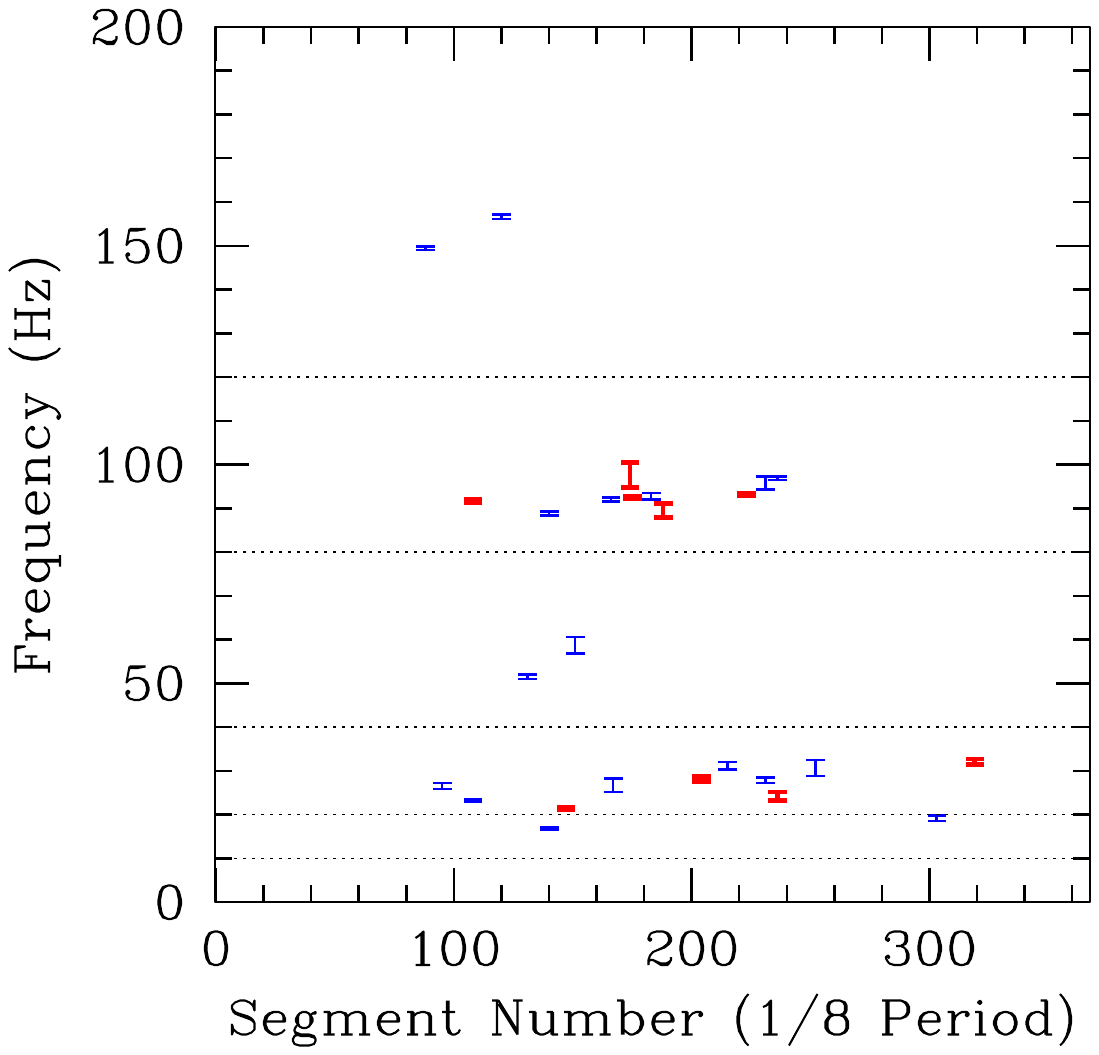}}
\vspace{-2.0truein}
   \caption{Plot of the best-fit centroid frequencies plus their $1\sigma$ uncertainties, for signals with Bayes factors ${\cal B}>1000$ and frequency widths less than 10~Hz.  The colors correspond to the magnitude of the Bayes factor: the thin blue bars are for $10^3<{\cal B}<10^5$, and the thick red bars are for ${\cal B}>10^5$.  The horizontal dotted lines show the boundaries between the frequency ranges that we search.}
\label{fig:strong}
\end{center}
\end{figure}

\begin{figure}
\begin{center}
  \resizebox{1.0\textwidth}{!}{\includegraphics{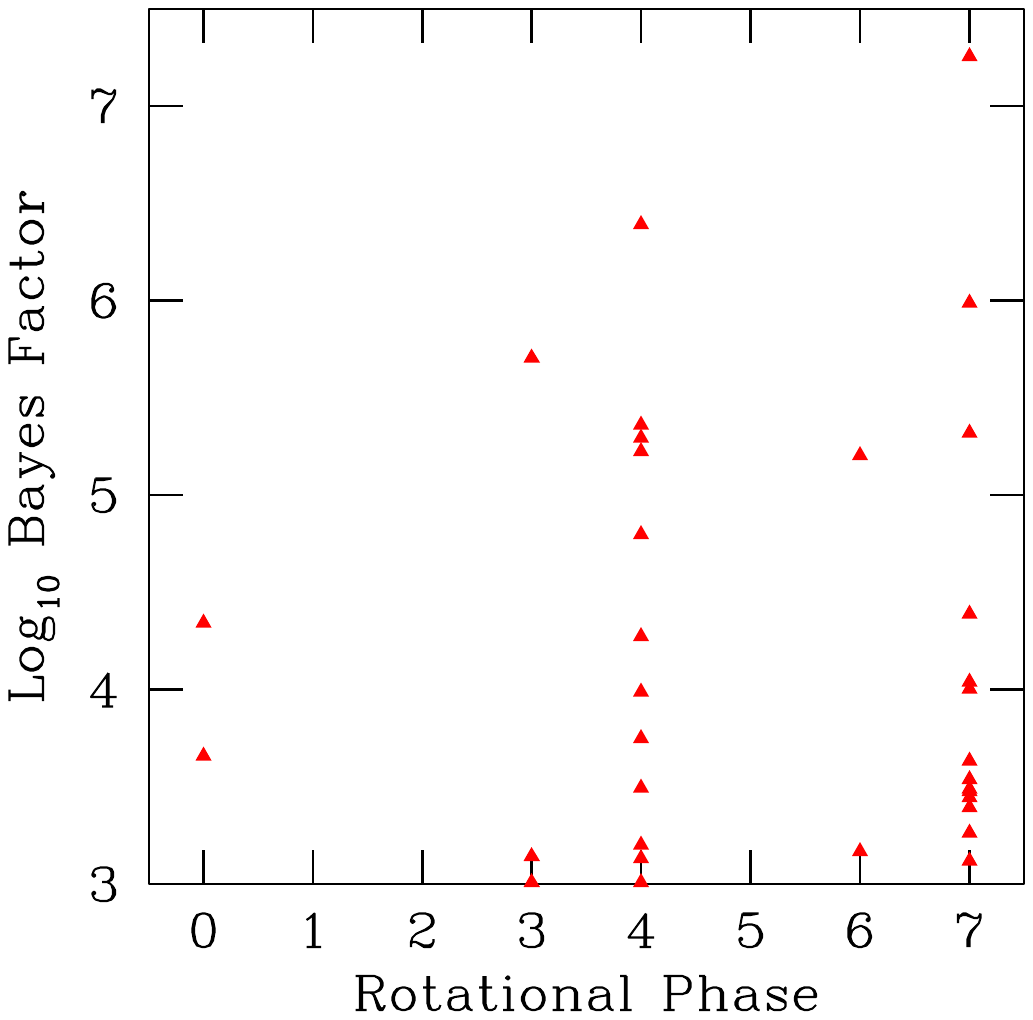}}
\vspace{-2.0truein}
   \caption{Rotational phases and log Bayes factors for all 1.0-second segments with Bayes factors greater than 1000 and frequency widths less than 10~Hz.  Each solid red triangle represents a separate signal.  Here the rotational phase refers to the eight segments of the period; see Figure~\ref{fig:lightcurve}.  The concentration in phases 4 and 7 is obvious; 26 of the 33 segments with a QPO that has a Bayes factor greater than 1000 are in those phases, and the remaining seven are all in adjacent phases.  This suggests that either the crust or the magnetosphere at those phases is particularly prone to the generation of the QPOs that we see.}
\label{fig:phases}
\end{center}
\end{figure}

In Figure~\ref{fig:strong} we show a summary of the ${\cal B}>1000$ segments with frequency width less than 10~Hz.  The error bars are centered on the best-fit centroid frequencies and their widths represent the 16\% to 84\% credibility range.  The color indicates the magnitude of the Bayes factor: the thin blue error bars are for $10^3<{\cal B}<10^5$, and the thick red error bars are for ${\cal B}>10^5$.

In Figure~\ref{fig:phases} we plot the rotational phases (as defined in Figure~\ref{fig:lightcurve}) at which QPOs with ${\cal B}>1000$ were detected.  The concentration at phases 4 and 7 is clear.  This suggests that something about those phases makes it particularly easy to see QPOs.  For example, it could be that the crust at those phases has properties that make modes especially easy to generate, or perhaps the magnetosphere at those phases facilitates the production of photons from modes.  The favored phases are not quite offset by half of the period, which makes it tempting to think that a slightly off-center dipole magnetic field could produce the favoritism in phase.

\subsection{Are there persistent signals?}

So far we have focused on short segments of data: 1.0 seconds, which is close to 1/8 of the rotational period.  Based just on the strengths of these signals as a function of segment number, there is no indication that the signals persist for more than one segment.  For example, our strongest signal has a Bayes factor of ${\cal B}=1.8\times 10^7$ at a frequency $f=32$~Hz.  In the $20-40$~Hz range, the previous segment has ${\cal B}=0.018$ and the following segment has ${\cal B}=0.432$.  Thus the duration of the signal is likely to be at most a few tenths of a second.  More generally, in the entire data set there is only one pair of consecutive 1-second segments with strong signals, and those signals have frequencies that do not overlap at the $2\sigma$ level.  The QPOs in this pair may therefore result from two completely independent excitations of modes.

There are, however, two additional possibilities to explore.  First, as suggested by \citet{2006ApJ...653..593S}, it could be that some modes persist for many rotational periods but are only visible during certain phases of the period.  Second, perhaps in addition to the quickly-damped oscillations that we detect with our 1.0-second analyses, there are weaker but longer-lasting oscillations at other frequencies.

To explore the first possibility we look at strong signals separated by eight segments, i.e., one full rotational period.  Because we want to look for signals that are also related to each other, we apply the following cuts: (1)~the signals must appear in consecutive periods rather than simply being separated by an integer number of periods, (2)~the $\pm 1\sigma$ centroid frequencies of the two periods must overlap, (3)~the frequency widths must be less than 10~Hz, and (4)~at least one of the segments must have a Bayes factor greater than 1000.  With these cuts, there are only two possible pairs: the segment starting at 203.175~seconds and the subsequent segment, where the frequency is $\approx 31$~Hz; and the segment starting at 165.375~seconds and the following segment, where the frequency is $\approx 92$~Hz.  In the second case, the following segment also has ${\cal B}>1000$.  

Thus in those two cases it is possible that there is an oscillation that persists for a full period.  However, given that the associated frequency ranges ($20-40$~Hz and $80-120$~Hz) are the sources of the most common, and the strongest, QPOs (see Figures~\ref{fig:bayes9panel} and \ref{fig:strong}), it also seems plausible that in both of these cases there were independent excitations, by happenstance, which were separated by one rotational period.  Some supporting evidence for this comes from the best-fit frequency widths, which are $\approx 2$~Hz for the 31~Hz signal and $\approx 0.7$~Hz for the 92~Hz signal.  There are many possible origins of the frequency widths, but if they are related to decay rates then over the 7.56-second rotational period we would expect the signals to drop to undetectability.

To explore the second possibility, of weaker persistent signals, we repeated our analysis for 8.0-second (roughly one period) segments that start every 7.56~seconds, and for 32.0-second (roughly four periods) segments that start every $4\times 7.56=30.24$~seconds (again, we choose to analyze data segments with durations that are an integer power of two larger than the $2^{-20}$ second time resolution in the {\it RXTE} data).  We are again looking for strong signals that have widths consistent with a damping time that is not much less than the duration of the segment.

From our 8.0-second analysis, our three best candidates are:

\begin{enumerate}

\item The segment starting 309.96~seconds from our zero point has a best-fit frequency of 20.307~Hz, a best-fit Lorentzian width of 0.1884~Hz, and a Bayes factor of 13577.1 relative to the pure red noise model.

\item The segment starting 83.16~seconds from our zero point has a best-fit frequency of 149.85~Hz, a best-fit Lorentzian width of 0.208~Hz, and a Bayes factor of 9097.86 relative to the pure red noise model.

\item The segment starting 60.48~seconds from our zero point has a best-fit frequency of 240.547~Hz, a best-fit Lorentzian width of 0.4529~Hz, and a Bayes factor of 21350.8 relative to the pure red noise model.  We note that no QPO at this frequency has been reported previously for these data.

\end{enumerate}

From our 32.0-second analysis, we have no good candidates; all four of the segments that give Bayes factors that are formally larger than 1000 have best-fit frequency widths that are several times larger than would be consistent with a persistent signal.

We therefore conclude that almost all of the QPOs are produced by events that last a few tenths of a second, and that at most a few QPOs can last as much as a few seconds.  Why, then, has it been possible for searches over much longer time scales to find signals?

We can provide some insight using a toy model.  Suppose that in a given data set there are $m$ independently excited flares that take the form of exponentially decaying periodic functions of angular frequency $\omega_0$; for flare $j$, which starts at time $t_j$, we represent the signal by a count rate
\begin{equation}
c_j(t)=a_je^{-\gamma_j(t-t_j)}e^{i\omega_0t}e^{i\phi_j}\; ,
\end{equation}
where $\phi_j$ is the phase.  The power density of the total signal, which we assume has a duration $T\gg 1/\gamma_j$, is
\begin{equation}
P(\omega)\propto\biggl|\int_0^T\sum_{j=1}^m c_j(t)e^{-i\omega t}dt\biggr|^2\; .
\end{equation}
If we assume for simplicity that $\gamma_j=\gamma$ and $a_j=a$ for all $j$, then the power density becomes
\begin{equation}
P(\omega)\propto {a^2\over{\gamma^2+(\omega-\omega_0)^2}}\biggl|\sum_{j=1}^m e^{i\phi_j}\biggr|^2\; .
\end{equation}
If all of the flares are in phase ($\phi_j=\phi$ for all $j$), then the squared factor is $m^2$; if the flares occur at random phases then the expectation value of the squared factor is $m$.

This indicates that even if $T\gg 1/\gamma$, the characteristic width of the peak in the power density spectrum will be $\sim\gamma$.  This provides a possible explanation for a puzzle noted by \citet{2006ApJ...653..593S}: that if the frequency widths are caused by exponential decay then the implied decay time would be far shorter than the lengths of the data segments being analyzed.

We also note that there are circumstances in which the signal could appear to be extremely significant over a relatively long interval of data.  If, for example, there are several flares whose phases happen to line up, then the power density will be increased significantly compared to the power density when the flares have phases that are more evenly distributed between 0 and $2\pi$.  Moreover, because power density spectra have greater frequency resolution when the interval is longer, even a weak signal with a significant frequency breadth could be detected with high significance because for long-duration data sets there will be many independent frequency bins with powers in excess of the Poisson average.  Thus it seems possible that most of the apparently persistent QPOs in the giant flare of SGR~1806$-$20 are actually composed of multiple independent flares.

\subsection{Can the frequencies be identified?}
\label{sec:freq_id}

We now compare our results for the strongest signals given in Table \ref{tab:QPOs} with physical models that could explain the process by which they were generated. If a QPO results from damped oscillations of the generic form
\begin{equation}
\psi(t) = Ae^{-\omega_{\rm I}t}\sin(\omega_{\rm R}t+\delta)\; ,
\end{equation}
then its power density is given by
\begin{equation}
P_{\rm QPO}(\omega) = |\Psi(\omega)|^2 = A^2\left(\frac{(\omega_{\rm R}\cos\delta + \omega_{\rm I}\sin\delta)^2 + \omega^2\sin^2\delta}{(\omega_{\rm R}^2 + \omega_{\rm I}^2 - \omega^2)^2 + 4\omega^2\omega_{\rm I}^2}\right)\; .
\label{eq:Ps_general}
\end{equation}
We stress that this is \emph{not} the same as the Lorentzian power density given in the second term of our Lorentzian model (Equation~\ref{eq:Lorentz}). In particular, the power density also depends on the phase $\delta$. From Figure~\ref{fig:strong} we see that there are several signals with close but distinct frequencies. Could it be  that they actually represent the same modes, only differing in phase?

It is possible to identify the parameters $f_0$ and $\Delta f$ of the phenomenological Lorentzian model with the oscillation frequency $\omega_{\rm R}$ and damping time $\tau \equiv 1/\omega_{\rm I}$ of the damped oscillation. Choosing as limiting cases $\delta = 0$  and $\delta = \pi/2$ we have, to leading order, 
\begin{eqnarray}
f_0\left[1-\frac{1}{2}\left(\frac{\Delta f}{f_0}\right)^2\right] < \frac{\omega_{\rm R}}{2\pi} < f_0\left[1+\frac{1}{2}\left(\frac{\Delta f}{f_0}\right)^2\right] \,,\\
\Delta f\left[1-\frac{1}{8}\left(\frac{\Delta f}{f_0}\right)^2\right] < \tau^{-1} < \Delta f\left[1+\frac{1}{8}\left(\frac{\Delta f}{f_0}\right)^2\right]\,. 
\end{eqnarray}
Consequently, for the strongest narrow QPOs reported in table \ref{tab:QPOs}, where we find a largest value of $\Delta f/f_0 = 0.17$, the maximum fractional correction is $\pm 0.015$  for the frequency and $\pm 0.004$ for the damping time. Therefore we can safely assume $\omega_{\rm R}/2\pi \approx f_0$ and $\tau^{-1} \approx \Delta f$. This indicates that a possible difference in phase is not enough to explain the variation in the values of the frequencies.

Another possibility for the variation in the frequencies is a time evolution of the magnetic field strength or geometry after the giant flare. However, this should cause a clearer trend in the frequency behavior than we see. Moreover, a variation of a few Hz in the frequency of the mode would require a change of several times $10^{15}$ G in the magnetic field (\citealt{1998ApJ...498L..45D}), which seems unreasonable. 

\begin{figure}
\begin{center}
  \resizebox{1.0\textwidth}{!}{\includegraphics{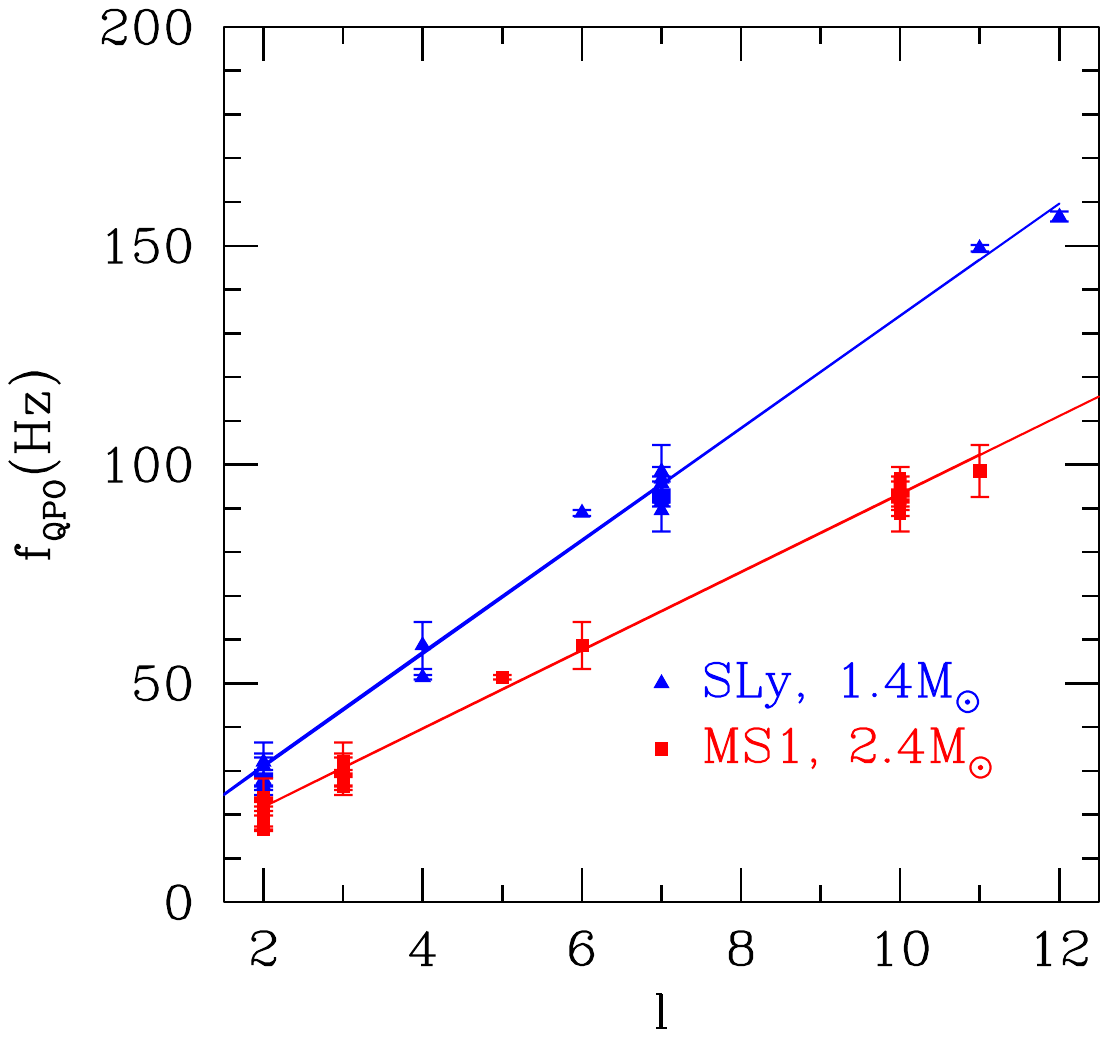}}
\vspace{-2.0truein}
   \caption{Frequency of the torsional modes of the crust of a neutron star as a function of the harmonic number $\ell$. The exact value of the frequency for each mode depends on the details of the structure and composition of the star. The two limiting cases presented here correspond to a $2.4~M_\odot$ with the MS1 equation of state (red line) and to a $1.4~M_\odot$ star with the SLy equation of state (blue line), where we used data obtained by \cite{deSouza2017}. The points show the strongest QPO frequencies found in our analysis and the corresponding nearest integer value of $\ell$ according to each model.}
\label{fig:fitlint}
\end{center}
\end{figure}

Using a simple model to describe the QPOs as torsional oscillations of the crust, it can be found that the frequencies of modes with different harmonic number $\ell$ should scale approximately as $f_{\ell} = f_0\sqrt{\ell(\ell+1)}$ (\citealt{1980ApJ...238..740H}). In Figure \ref{fig:fitlint} we choose two limiting values of $f_0$: one corresponding to a $2.4~M_\odot$ star with the MS1 equation of state, and the other corresponding to a $1.4~M_\odot$ star with the SLy equation of state.  The expected frequencies in these two models, as a function of the harmonic number $\ell$, are shown by the two solid lines. We use the resulting expressions to find the corresponding nearest integer value of $\ell$ for our frequencies, which are shown with points. As the plot shows, the frequencies we found with our analysis of the data are compatible with torsional oscillations of the crust, but the mode identification would depend on the largely unknown features of the star.  Although the mode identification is uncertain, our results are compatible with previous findings that show that not every $\ell$ seems to be strongly excited.  This could be a consequence of the initial perturbation, i.e., the exact manner in which the crust was originally broken.

\section{DISCUSSION}
\label{sec:discussion}

Now we explore the implications our results could have for a model of the source star, the nature of the oscillations, and the mechanism responsible for the emission, that is, in what way the stellar oscillations can be coupled to the radiation emitted by the star and observed on Earth.

%Implications for magnetic field geometry?
%Considerations of emission (e.g., beamed versus broad)?

The results presented in Figure~\ref{fig:phases} show that the emission happened predominantly in two nearly opposite rotational phases labeled 4 and 7 in our notation; one full period goes from phases 0 to 7 in our analysis).
\footnote{It is also worth noting that this phase dependence in the appearance of the QPOs is distinct from the phase dependence observed in the number of counts (see Figure~\ref{fig:lightcurve}), indicating that it should be a real effect, and not a bias resulting from the detection of more or less counts.}
A possible way of obtaining this symmetry would be by means of a slightly off-center dipolar magnetic field. The physical mechanism for the emission is not yet clearly understood (but see the models proposed by \citealt{2008ApJ...680.1398T,2012ApJ...751L..41D,2014AN....335..240G}).  Whether it comes from the crust or from the magnetosphere, the QPOs appear to be strong mostly in those two phases. If the giant flare indeed comes from a rearrangement of the magnetic field, the crust at the magnetic poles of the star could be the regions more likely to break; at the same time, the magnetosphere above the poles would have the largest magnetic energy density. Both conditions could amplify the signal in a way such that it would be mostly visible at these two phases, making it seem more beamed, even if it is a broad (thermal) emission. 
%
%This picture is consistent with the models proposed by \cite{2008ApJ...680.1398T}, who propose that the emission originates at the magnetosphere and is caused by crustal oscillations that shake the magnetic field lines\ldots \cite{2012ApJ...751L..41D}\ldots \cite{2014MNRAS.441.2676L}\ldots \cite{2014AN....335..240G}\ldots (ADD DETAILS).

%Implications for MHD coupling and damping?

Perhaps our most striking result is the indication that there are no obviously persistent oscillations in the tail of the giant flare.  This is expected from theoretical studies of the properties of the crustal oscillations, which indicate that the crustal modes will couple to a continuum of MHD modes excited below the crust, which quickly damp the crustal oscillations. If, as we discussed in Section \ref{sec:freq_id}, we estimate the damping time $\tau$ of the oscillations by the inverse of the frequency width $\Delta f$, our results in Table~\ref{tab:QPOs} imply $\tau \approx 0.2 - 2$ s. This is consistent with the findings of \cite{2014ApJ...793..129H} for a higher frequency QPO ($\tau \approx 0.5$ s), and also roughly consistent with the expectations of \cite{2006MNRAS.368L..35L} ($\tau$ at most 1 s), which takes into account the damping resulting from the coupling with the continuum spectrum of MHD modes. 

More sophisticated theoretical analyses have introduced the possibility that there are gaps in the continuous spectrum of MHD modes, which could allow for longer-lived oscillations at least for the lower frequencies detected in the QPOs. Gaps in the continuum could be a consequence of more complicated magnetic field geometries (for example including a toroidal component or a tangled magnetic field: see \cite{2016ApJ...823L...1L,2016ApJS..224....6L}). However, our analysis shows no compelling evidence for persistent oscillations in any part of the spectrum.  This therefore suggests that the QPOs are independent oscillations with distinct but close frequencies (see Figure~\ref{fig:strong}). This could be understood as evidence of the existence of the continuous spectrum of MHD modes that was theoretically predicted. If the models for the continuum gap are correct, then the lack of persistent oscillations provides further support for a simple magnetic field configuration close to a pure dipole. 

Another consequence of this picture is the need for a continued re-excitation of the modes after the giant flare. If the initial shock causes a starquake, then perhaps aftershocks in the crust provide the energy for the subsequent excitations. This could provide constraints on the nature of the crust, but other unknown mechanisms, perhaps including interactions with the perturbed magnetosphere, could be responsible for the continued input of energy.

Finally, we have not attempted to perform an identification of the modes we have obtained in our analysis. Even in the simplest scenario in which the QPOs are explained as torsional crustal oscillations, the exact frequencies will depend on the mass, compactness, equation of state, crust composition and shear modulus, magnetic field strength and geometry, and so on. This multitude of parameters makes it extremely challenging to identify the modes and to solve the inverse problem, particularly given that there is a lack of relevant analytical expressions or universal relations for these frequencies (work on these issues is in preparation by G. de Souza and C. Chirenti). Nonetheless, the detailed data available for these QPOs means that they can still serve as a rich source of information that can be used to constrain many aspects of the interiors of neutron stars.

\acknowledgements

\noindent {\bf Acknowledgements}

We thank Fred Lamb, Yuri Levin, and Bennett Link for valuable discussions and comments on an earlier version of this manuscript.  This work was supported in part by joint research workshop award 2015/50421-8 from FAPESP and the University of Maryland, College Park.

\bibliography{ms}

\end{document}